\documentclass[english]{article}
\usepackage[T1]{fontenc}
\usepackage[latin9]{luainputenc}
\usepackage{geometry}
\usepackage{amstext}
\usepackage{amssymb}
\usepackage{feyn}
\usepackage{tikz}
\usepackage{graphicx}
\usepackage{array}
\usepackage{booktabs}

\usepackage{amsmath}
\usepackage{amsfonts}

\usepackage[normalem]{ulem}
\usepackage{color}
\usepackage{listings,braket}
\usepackage{caption}
\usepackage{subcaption}
\usepackage{float}
\definecolor{darkgreen}{rgb}{0,0.35,0}
\definecolor{Rood}{rgb}{1, 0, 0}

\makeatletter
\usepackage[english]{babel}
\usepackage{feyn}

\makeatother

\begin{document}

\title{\textbf{Entanglement and maximal violation of the CHSH inequality in a system of two spins  $j$: a novel  construction and further observations}}

{\author{\textbf{G.~Peruzzo$^1$}\thanks{gperuzzofisica@gmail.com},
		\textbf{S.~P.~Sorella$^2$}\thanks{silvio.sorella@gmail.com},\\\\\
		\textit{{\small $^1$Instituto de F\'{i}sica, Universidade Federal Fluminense,
				}}\\
		\textit{{\small Campus da Praia Vermelha, Av. Litor\^{a}nea s/n, }}\\
		\textit{{\small 24210-346, Niter\'{o}i, RJ, Brasil}}\\
		\textit{{\small $^2$UERJ -- Universidade do Estado do Rio de Janeiro,}}\\
		\textit{{\small Instituto de F\'{\i}sica -- Departamento de F\'{\i}sica Te\'orica -- Rua S\~ao Francisco Xavier 524,}}\\
		\textit{{\small 20550-013, Maracan\~a, Rio de Janeiro, Brasil}}\\
	}

\date{}

\maketitle
\begin{abstract}
	We study the CHSH inequality for a system of two spin $j$ particles, for generic $j$. 	  The CHSH operator is constructed using a set of unitary, Hermitian operators $\left\{ A_{1},A_{2},B_{1},B_{2}\right\} $. The expectation value of the CHSH operator is analyzed for the singlet state $\left|\psi_{s}\right\rangle $. Being 
 $\left|\psi_{s}\right\rangle $  an entangled state, a violation of the CHSH inequality compatible with Tsirelson's
bound is found. Although the construction employed here differs from that of \cite{Peres1}, full agreement is recovered. 
\end{abstract}

\section{Introduction}

In Quantum Mechanics, the general absence of an objective reality prior
to a measurement leads to conflicts with the principles of relativity,
as locality. For instance, if we have a composed system and, according
to Quantum Mechanics, its observable quantities are not defined before
measuring them, a measurement done in  some part of the system necessarily
 affects instantaneously  the other parts, so as to  keep the
very strong correlations which arise due to the phenomenon of entanglement. 
This somehow nonlocal feature of Quantum Mechanics was used by Einstein-Podolsky-Rosen
(EPR) \cite{Einstein:1935rr} to argue that Quantum Mechanics was an incomplete theory, despite of being logically sound.

Many physicists, as  \cite{Bohm:1951xw}, tried to supplement the Quantum Mechanics framework
with a set of local hidden variables which, once known,  would provide a complete deterministic description as opposed to a probabilistic one. This long debate between supporters of Quantum Mechanics and supporters
of what  is usually called local realism received  a pivotal  contribution from
John S. Bell \cite{Bell:1964kc,Bell:1964fg}, who found, through Bell's inequalities, a way
to test both points of view. One of the beauties of Bell's work is
its generality that allows us to test, in principle, the local realism 
in any instance. Afterwards, Clauser-Horne-Shimony-Holt
(CHSH) \cite{Clauser:1969ny,Freedman:1972zza,Clauser:1974tg,Clauser:1978ng} proposed a generalized version of Bell's idea more
suitable for direct experiments. A violation of Bell's inequalities or CHSH
inequalities means that nature cannot be described by a local realistic
model. In fact, Quantum Mechanics predicts a violation of Bell's
inequalities or CHSH inequalities in, basically, any physical entangled system,
like the two-particles system used by EPR to refute Quantum Mechanics.
Today, despite the high sophistication needed to test Bell's or
CHSH inequalities experimentally,  see, for instance,  \cite{Freedman:1972zza,aspect0,aspect1,aspect2,aspect3,z1},  
Quantum Mechanics remains a solid,  robust and consistent  framework to handle the quantum world. \\\\As the two-spin systems are the simplest systems to study entanglement, we use them as an ideal laboratory to investigate how the maximal violation of the CHSH inequality behaves with respect to the values of the spin $j$.\\\\More specifically, we introduce in a system of two spins
$j$ a set of unitary, Hermitian operators, $\left\{ A_{1},A_{2},B_{1},B_{2}\right\} $,
compatible with the CHSH inequality. Using these operators, we construct
the CHSH operator, $\mathcal{O}_{CHSH}$, and analyze its expectation
value in the entangled singlet state $\left|\psi_{s}\right\rangle $.  Computing
the expectation value of $\mathcal{O}_{CHSH}$,
$\left\langle \psi_{s}\right|\mathcal{O}_{CHSH}\left|\psi_{s}\right\rangle $,
we get an expression that depends on $j$ and, analyzing it, we notice
a difference between $j$ integer and $j$ half-integer. In both cases,
 Tsirelson's bound \cite{tsi1,tsi2,tsi3} for the violation of CHSH inequality
is satisfied, being saturated in the half-integer case. \\\\It is worth underlining that the construction of the operators $\left\{ A_{1},A_{2},B_{1},B_{2}\right\} $ is different from that of \cite{Peres1}. As outlined in 
\cite{Sorella:2023pzc}, in addition of its simplicity, the setup presented here has rather general applicability, covering models ranging from Quantum Mechanics to relativistic Quantum Field Theory \cite{Sorella:2023pzc}. Though, full agreement with  the general results established by \cite{Peres1} is achieved.

 .

\section{Construction  of the operators $A_i$ and $B_i$}

Let us consider a two spins $j$ system. The Hilbert space,
$\mathcal{H}$, is spanned by the tensor product of the orthonormal
states $\left\{ \left|m\right\rangle ;\,-j\leq m\leq j\right\} $,
that is by the set
\begin{eqnarray}
&  & \left\{ \left|m\right\rangle \otimes\left|n\right\rangle ;\,-j\leq m,n\leq j\right\} .\label{eq:tensor_basis}
\end{eqnarray}
which means that ${\rm dimension} \;\mathcal{H}=\left(2j+1\right)^{2}$ . In order to construct a CHSH operator we define
a set of unitary, Hermitian operators
\begin{eqnarray}
\left\{ A_{1},\,A_{2},\,B_{1},\,B_{2}\right\} \label{eq:set_of_operators}
\end{eqnarray}
by their action on the basis \eqref{eq:tensor_basis}:
\begin{eqnarray}
A_{i}\left|-m\right\rangle \otimes\left|n\right\rangle  & = & e^{i\left(\alpha_{i}\right)_{m}}\left|m\right\rangle \otimes\left|n\right\rangle ,\quad\left(\alpha_{i}\right)_{m}\in\mathbb{R},\quad\left(\alpha_{i}\right)_{-m}=-\left(\alpha_{i}\right)_{m}\label{eq:A_definition}
\end{eqnarray}
and
\begin{eqnarray}
B_{i}\left|m\right\rangle \otimes\left|-n\right\rangle  & = & e^{-i\left(\beta_{i}\right)_{n}}\left|m\right\rangle \otimes\left|n\right\rangle ,\quad\left(\beta_{i}\right)_{n}\in\mathbb{R},\quad\left(\beta_{i}\right)_{-n}=-\left(\beta_{i}\right)_{n}\label{eq:B_definition}
\end{eqnarray}
for all$-j\leq m,n\leq j$. \\\\Note that $A_i$ and $B_i$ change only one state of the tensor product, which means that they could have been defined as $A_i \otimes I$ and $I \otimes B_i$. One can easily check that $A_{i}$
and $B_{i}$ are unitary, Hermitian operators, \emph{i.e. }
\begin{eqnarray}
A_{i}^{\dagger} & = & A_{i},\nonumber \\
B_{i}^{\dagger} & = & B_{i},
\end{eqnarray}
and
\begin{eqnarray}
A_{i}^{2} & = & I,\nonumber \\
B_{i}^{2} & = & I.
\end{eqnarray}
Moreover, from the definitions \eqref{eq:A_definition} and \eqref{eq:B_definition}
it follows that
\begin{equation}
	\left[A_{i},B_{j}\right]=0.
\end{equation}
Each operator, $A_{i}$ or $B_{j}$, is characterized by a set of
real parameters, $\alpha_{i}\equiv\left\{ \left(\alpha_{i}\right)_{m};\,-j\leq m\leq j\right\} $
or $\beta\equiv\left\{ \left(\beta_{i}\right)_{m};\,-j\leq m\leq j\right\} $,
being $\alpha_{i}$ and $\beta_{j}$ completely independent.
This feature makes $A_{i}$ and $B_{j}$ compatible with one of the main requirements
used to derive Bell's inequality: \emph{the experimental arrangement
to measure $A_{i}$ is totally independent from the  arrangement
to measure $B_{j}$}. This requirement is important because it allows
to make two causal, {\it i.e.} space-like separated independent measurements of $A_{i}$ and $B_{j}$. Following Bell's methodology,
it is then possible to test in this model if the results predicted
by the Quantum Mechanics formalism are compatible with some \emph{local
	realistic theory},  like a theory based on  local hidden variables. 

\section{Expectation value of the CHSH operator in the singlet state }

We introduce the CHSH operator, $\mathcal{O}_{CHSH}$, in its traditional
form, namely 
\begin{eqnarray}
\mathcal{O}_{CHSH} & = & \left(A_{1}+A_{2}\right)B_{1}+\left(A_{1}-A_{2}\right)B_{2}.\label{eq:chsh_operator}
\end{eqnarray}
By construction,  $\mathcal{O}_{CHSH}$ is  Hermitian,  $\mathcal{O}_{CHSH}^{\dagger}=\mathcal{O}_{CHSH}$. \\\\From the point of view of the local realism, one would expect 
\begin{eqnarray}
-2\leq & \left\langle \mathcal{O}_{CHSH}\right\rangle  & \leq2.\label{eq:chsh_inequalities}
\end{eqnarray}
Relation \eqref{eq:chsh_inequalities} is  the Bell-Clauser-Horne-Shimony-Holt
inequality. Now, we evaluate the expectation value of \eqref{eq:chsh_operator}
in the singlet state $\left|\psi_{s}\right\rangle $, defined as \footnote{By singlet we mean the state $\left|\psi_{s}\right\rangle $ such
	that
	\begin{eqnarray}
	S_{x}\left|\psi_{s}\right\rangle  & = & 0,\nonumber \\
	S_{y}\left|\psi_{s}\right\rangle  & = & 0,\nonumber \\
	S_{z}\left|\psi_{s}\right\rangle  & = & 0,\label{eq:singlet_properties}
	\end{eqnarray}
	where $\vec{S}=\vec{S}_{1}+\vec{S}_{2}$ is the total spin. As a consequence
	of \eqref{eq:singlet_properties}, $\left|\psi_{s}\right\rangle $ is
	rotational invariant. }
\begin{eqnarray}
\left|\psi_{s}\right\rangle  & = & \frac{1}{\left(2j+1\right)^{\frac{1}{2}}}\sum_{m=-j}^{j}\left(-1\right)^{j-m}\left|m\right\rangle \otimes\left|-m\right\rangle .\label{eq:singlet-state}
\end{eqnarray}
One can check that $\left|\psi_{s}\right\rangle $ is a normalized
state, $\left\langle \psi_{s}|\psi_{s}\right\rangle =1$. Since
\begin{eqnarray}
	A_{i}B_{j}\left|\psi_{s}\right\rangle  & = & \frac{1}{\left(2j+1\right)^{\frac{1}{2}}}\sum_{m=-j}^{j}\left(-1\right)^{j-m}e^{-i\left(\left(\alpha_{i}\right)_{m}+\left(\beta_{j}\right)_{m}\right)}\left|-m\right\rangle \otimes\left|m\right\rangle ,
\end{eqnarray}
we have the following expectation value
\begin{eqnarray}
\left\langle \psi_{s}\right|A_{i}B_{j}\left|\psi_{s}\right\rangle  & = & \frac{1}{\left(2j+1\right)}\sum_{n=-j}^{j}\sum_{m=-j}^{j}\left(-1\right)^{j-n}\left(-1\right)^{j-m}e^{-i\left(\left(\alpha_{i}\right)_{m}+\left(\beta_{j}\right)_{m}\right)}\delta_{n,-m}\delta_{-n,m}\nonumber \\
& = & \frac{\left(-1\right)^{2j}}{\left(2j+1\right)}\sum_{m=-j}^{j}e^{-i\left(\left(\alpha_{i}\right)_{m}+\left(\beta_{j}\right)_{m}\right)}.
\end{eqnarray}
Therefore, we get for the CHSH operator the following result
\begin{eqnarray}
	\left\langle \psi_{s}\right|\mathcal{O}_{CHSH}\left|\psi_{s}\right\rangle  & = & \frac{\left(-1\right)^{2j}}{\left(2j+1\right)}\sum_{m=-j}^{j}\left[e^{-i\left(\left(\alpha_{1}\right)_{m}+\left(\beta_{1}\right)_{m}\right)}+e^{-i\left(\left(\alpha_{2}\right)_{m}+\left(\beta_{1}\right)_{m}\right)}+e^{-i\left(\left(\alpha_{1}\right)_{m}+\left(\beta_{2}\right)_{m}\right)}-e^{-i\left(\left(\alpha_{2}\right)_{m}+\left(\beta_{2}\right)_{m}\right)}\right] \nonumber \\
	& & 
\end{eqnarray}
It is useful now to distinguish the two cases:
\begin{itemize}
	\item $j\geq1$ integer 
\end{itemize}
\begin{eqnarray}
	\left\langle \psi_{s}\right|\mathcal{O}_{CHSH}\left|\psi_{s}\right\rangle  & = & \frac{1}{\left(2j+1\right)}\left\{ 2+2\sum_{m=1}^{j}\left[\cos\left(\left(\alpha_{1}\right)_{m}+\left(\beta_{1}\right)_{m}\right)+\cos\left(\left(\alpha_{2}\right)_{m}+\left(\beta_{1}\right)_{m}\right)\right.\right.\nonumber \\
	&  & \left.\left.+\cos\left(\left(\alpha_{1}\right)_{m}+\left(\beta_{2}\right)_{m}\right)-\cos\left(\left(\alpha_{2}\right)_{m}+\left(\beta_{2}\right)_{m}\right)\right]\right\} 
\end{eqnarray}
We have that
\begin{eqnarray}
-2\sqrt{2}\leq\cos\left(\left(\alpha_{1}\right)_{m}+\left(\beta_{1}\right)_{m}\right)+\cos\left(\left(\alpha_{2}\right)_{m}+\left(\beta_{1}\right)_{m}\right)+\cos\left(\left(\alpha_{1}\right)_{m}+\left(\beta_{2}\right)_{m}\right)-\cos\left(\left(\alpha_{2}\right)_{m}+\left(\beta_{2}\right)_{m}\right) & \leq & 2\sqrt{2}\label{eq:image_cos_function} \nonumber \\
& &
\end{eqnarray}
for any $m$. If we choose, for example, the values 
\begin{eqnarray}
\left(\alpha_{1}\right)_{m} & =-\frac{\pi}{4} & ,\quad\left(\alpha_{2}\right)_{m}=\frac{\pi}{4},\quad\left(\beta_{1}\right)_{m}=0,\quad\left(\beta_{2}\right)_{m}=\frac{\pi}{2},\label{eq:set_parameters}
\end{eqnarray}
then we get the maximal violation
\begin{eqnarray}
	  \left|\left\langle \psi_{s}\right|\mathcal{O}_{CHSH}\left|\psi_{s}\right\rangle \right| & = & \frac{2}{\left(2j+1\right)}\left(1+2j\sqrt{2}\right) > 2 \label{eq:maximal_violation_integer}
\end{eqnarray}
Since $\frac{2}{\left(2j+1\right)}\left(1+2j\sqrt{2}\right)$ is a
monotonic nondecreasing function and $\lim_{j\rightarrow\infty}\frac{2}{\left(2j+1\right)}\left(1+2j\sqrt{2}\right)=2\sqrt{2}$,
it follows that
\begin{equation}
	2 < \left|\left\langle \psi_{s}\right|\mathcal{O}_{CHSH}\left|\psi_{s}\right\rangle \right|  <  2\sqrt{2},
\end{equation}
for all $j\geq1$ integer. 

\begin{itemize}
	\item $j$ half-integer
\end{itemize} 
\begin{eqnarray}
	\left\langle \psi_{s}\right|\mathcal{O}_{CHSH}\left|\psi_{s}\right\rangle  & = & -\frac{2}{\left(2j+1\right)}\sum_{m=\frac{1}{2}}^{j}\left[\cos\left(\left(\alpha_{1}\right)_{m}+\left(\beta_{1}\right)_{m}\right)+\cos\left(\left(\alpha_{2}\right)_{m}+\left(\beta_{1}\right)_{m}\right)\right.\nonumber\\
	&  & \left.+\cos\left(\left(\alpha_{1}\right)_{m}+\left(\beta_{2}\right)_{m}\right)-\cos\left(\left(\alpha_{2}\right)_{m}+\left(\beta_{2}\right)_{m}\right)\right]
\end{eqnarray}
So, in this case, according to \eqref{eq:image_cos_function}, we immediately
get
\begin{eqnarray}
	\left|\left\langle \psi_{s}\right|\mathcal{O}_{CHSH}\left|\psi_{s}\right\rangle \right| & \leq & 2\sqrt{2}
\end{eqnarray}
for all $j$ half-integer. Furthermore, for the particular choice \eqref{eq:set_parameters},
it follows that the maximal violation is attained, {\it i.e.}
\begin{eqnarray}
	\left|\left\langle \psi_{s}\right|\mathcal{O}_{CHSH}\left|\psi_{s}\right\rangle \right| & = & 2\sqrt{2} \label{eq:maximal_violation_half_integer}.
\end{eqnarray}
Note the difference between integer and half-integer spin.  In the
integer case the maximum violation, $2\sqrt{2} $, is never reached, whereas in the half-integer case is attained. We point out
another important fact: all  results are compatible with the
Tsirelson bound \cite{tsi1,tsi2,tsi3}, which establishes that the maximum
violation is exactly $2\sqrt{2}$. Also, we underline that results \eqref{eq:maximal_violation_integer} and \eqref{eq:maximal_violation_half_integer} agree with the results of \cite{Peres1}. In fact, our construction of the operators $\{ A_1,\,A_2,\,B_1,\,B_2 \}$, see Eqs.\eqref{eq:A_definition} and \eqref{eq:B_definition}, can be seen as a concrete example of the general result demonstrated by \cite{Peres1} that \emph{for any nonfactorable state of two quantum systems, like $\left|\psi_{s}\right\rangle $, it is possuble to find  pairs of observables whose correlations violate the Bell-Clauser-Horne-Shimony-Holt inequality}. 

\section{Conclusions}
In this work we have employed the setup outlined in \cite{Sorella:2023pzc} to construct the four operators $\{ A_1,\,A_2,\,B_1,\,B_2 \}$ entering the CHSH inequality in the case of a pair of spin $j$ particles. As pointed out in \cite{Sorella:2023pzc}, the procedure is quite simple, being of general applicability in many cases, ranging from Quantum Mechanics to relativistic Quantum Field Theory \cite{Sorella:2023pzc}. \\\\In particular, in the present case, the results of \cite{Peres1} have been re-obtained in a simple and efficient way. 

\section*{Acknowledgements}
The authors would like to thank the Brazilian agencies CNPq and FAPERJ for financial support. This study was financed in part
by the Coordena{\c c}{\~a}o de Aperfei{\c c}oamento de Pessoal de N{\'\i}vel Superior--Brasil (CAPES) --Finance Code 001. S.P.~Sorella is a level $1$ CNPq researcher under the contract 301030/2019-7.

\end{document}